\documentclass{svjour3}                     
\smartqed  
\usepackage{graphicx}
\journalname{Few-Body Systems (FB20)}
\begin{document}

\title{
Spectroscopy of \boldmath{$\eta '$} mesic nuclei with (\boldmath{$p$},\boldmath{$d$}) reaction
\thanks{Presented at the 20th International IUPAP Conference on Few-Body Problems in Physics, 20 - 25 August, 2012, Fukuoka, Japan}
}
\author{Yoshiki~K.~Tanaka\and
Stefan~Friedrich\and
Hiroyuki~Fujioka\and
Hans~Geissel\and
Ryugo~S.~Hayano\and
Satoru~Hirenzaki\and
Kenta~Itahashi\and
Satoshi~Itoh\and
Daisuke~Jido\and
Volker~Metag\and
Hideko~Nagahiro\and
Mariana~Nanova\and
Takahiro~Nishi\and
Kota~Okochi\and
Haruhiko~Outa\and
Ken~Suzuki\and
Takatoshi~Suzuki\and
Helmut~Weick}
\authorrunning{Short form of author list} 
\institute{Y. K. Tanaka, R. S. Hayano, S. Itoh, T. Nishi, K. Okochi, T. Suzuki \at
Department of Physics, University of Tokyo, 113-0033 Tokyo, Japan\\
Tel.: +81-3-5841-4236\\
Fax: +81-3-5841-7642\\
\email{tanaka@nucl.phys.s.u-tokyo.ac.jp}           
\and
K. Itahashi, H. Outa \at
RIKEN Nishina Center, RIKEN, 351-0198 Saitama, Japan 
\and 
H. Fujioka \at
Division of Physics and Astronomy, Kyoto University, 606-8502 Kyoto, Japan
\and 
H. Geissel, H. Weick \at
GSI - Helmholtzzentrum f\"{u}r Schwerionenforschung GmbH, D-64291 Darmstadt, Germany  
\and 
S. Friedrich, V. Metag, M. Nanova \at
II. Physikalisches Institut, Universit\"{a}t Gie{\ss}en, D-35392 Gie{\ss}en, Germany 
\and 
S. Hirenzaki, H. Nagahiro\at
Department of Physics, Nara Women's University, 630-8506 Nara, Japan    
\and 
D. Jido\at
Yukawa Institute for Theoretical Physics, Kyoto University, Kyoto, 606-8502, Japan   
\and 
K. Suzuki\at
Stefan-Meyer-Institut f\"{u}r subatomare Physik, \"{O}sterreichische Akademie der Wissenschaften, 1090 Vienna, Austria}

\date{Received: date / Accepted: date}

\maketitle

\begin{abstract}

We are going to perform an inclusive spectroscopy experiment 
of $\eta '$ mesic nuclei with the $^{12}$C($p$,$d$) reaction
to study in-medium properties of the $\eta '$ meson. 
In nuclear medium, the $\eta '$ meson mass may be reduced due to partial restoration of chiral symmetry. 
In case of sufficiently large mass reduction and small absorption width of $\eta '$ at normal nuclear density, 
peak structures of $\eta '$ mesic states in ${}^{11}$C 
will 
be observed near the $\eta '$ emission threshold even in an inclusive spectrum.
The experiment will be carried out at GSI with proton beam supplied by SIS using FRS as a spectrometer.
The detail of the experiment
is described.

\end{abstract}
\section{Introduction}
\label{intro}
One important feature of the $\eta '$ meson is its especially heavy mass compared to the other pseudoscalar mesons,  
which is theoretically understood as the effect of $U_A(1)$ anomaly.
The strength of this effect, pushing up its mass, is considered to be related to chiral condensate \cite{Ref1,Ref1_2}.
Then, in nuclear medium, due to partial restoration of chiral symmetry, the mass of $\eta '$ may be reduced. 
Actually, in the NJL model calculations, it is shown that 
the mass reduction accounts for about 150 MeV at normal nuclear density \cite{Ref2,Ref3}.
Such mass reduction serves as attractive potential between an $\eta '$ and a nucleus. Therefore, $\eta '$ meson-nucleus bound states may exist \cite{Ref3,Ref4}. 

So far, there is no direct experimental information on $\eta '$ mass in medium, but scattering length of $\eta '$-nucleon interaction, which is related to the mass reduction.
In Ref.~\cite{Ref5}, from measurements of $pp \to pp\eta '$ reaction near threshold, scattering length of $\eta '$-nucleon interaction was evaluated to be of the order of 0.1 fm.
This suggests that interaction between $\eta '$ meson and nucleon is not strong, and it seems difficult to understand the small scattering length and the scenario of large mass reduction
at the same time \cite{Ref4}. In this sense, a new experiment on $\eta '$ mesic nucleus states will give new information on this situation.

As for the in-medium width, the CBELSA/TAPS collaboration found small absorption width of $\eta '$ at normal nuclear density \cite{Ref6}.
They measured mass-number dependence of the transparency ratio, and concluded $\eta '$ absorption 
width of 15-25 MeV at normal nuclear density for the average $\eta '$ momentum of 1050 MeV/$c$.
This suggests that decay width of $\eta '$ meson-nucleus bound states could be small as well.
Therefore, one may observe $\eta '$-nucleus bound states as distinct peaks experimentally.
\section{Experimental Method}
\label{exp}

\begin{figure}[b]
  \includegraphics[scale=0.4]{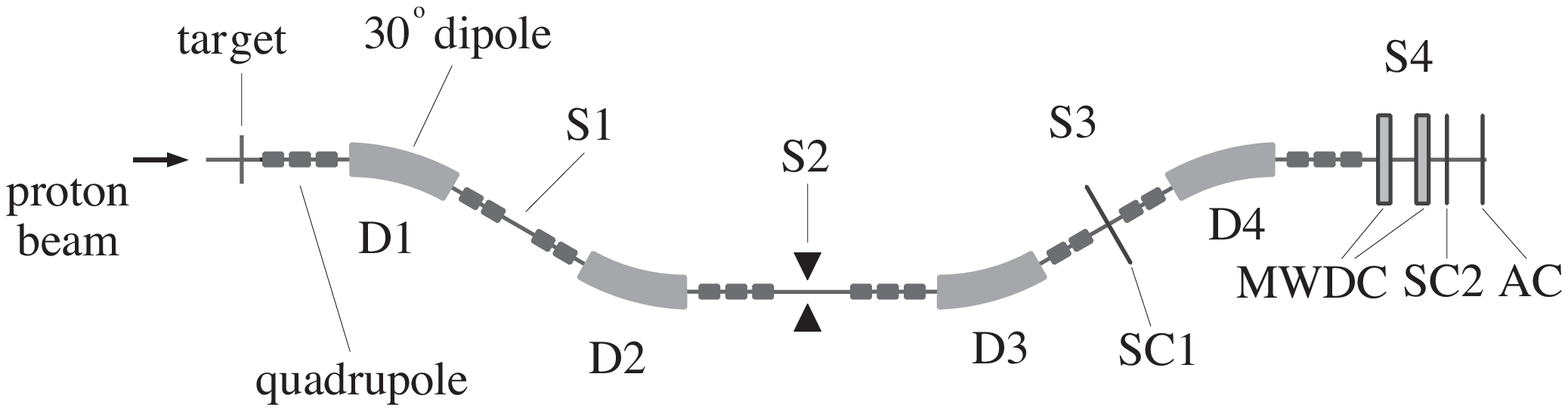}
\caption{Schematic view of the setup in FRS. See the text for the detail.}
\label{fig_frs_setup}       
\end{figure}

We are planning a missing mass spectroscopy of the $^{12}$C($p$,$d$) reaction near the $\eta '$ emission threshold at GSI \cite{Ref7,Ref8}.
We will employ 2.5 GeV proton beam from SIS (Heavy Ion Synchrotron) with the rate of $10^{10}/$s, and use 4 g/cm${}^2$-thick carbon as a target. 
In order to measure the momentum of outgoing deuterons, we will use FRS (Fragment Separator) as a forward spectrometer.
Figure \ref{fig_frs_setup} shows our setup in FRS. 
We will install two sets of multiwire drift chambers (MWDC) at a dispersive focal plane at S4.
By measuring positions of ejectile deuterons, missing masses in the reaction can be obtained.
Owing to good resolution of FRS, the overall spectral resolution will be about 1.6 MeV, which is sufficiently small compared to the decay width expected. 

In the experiment, however, also many background particles can reach the S4 area. 
Firstly, intense primary beam will be dumped in the first dipole magnet, and produce many secondary particles. 
This background can be suppressed by adopting an appropriate optics mode. We adopt momentum-compaction 
optics at the middle focal plane S2 and install slits there 
in order to select only particles originating from the target and suppress such secondary background. 
Secondly, we expect proton background produced by the ($p$,$p'$) reaction in the target. 
To reject this proton background, we will install an aerogel Cherenkov counter (AC) at S4 and scintillation counters (SC1, SC2) at S3 and S4.
In the trigger level, most of the protons will be rejected by the aerogel Cherenkov counter, which has
a threshold between the velocity of the signal deuterons and that of the background protons. 
In the off-line analysis, by use of time-of-flight between S3 and S4, we expect almost all these background can be rejected.

One feature of this experiment is an inclusive spectroscopy, in which only ejectile deuterons are measured.
This leads to a simple analysis, as no assumption on decay processes is necessary.
However, the signal-to-noise ratio becomes poor, because of 
quasi-free meson (not $\eta'$) production processes
as described in the next section. This can be overcome by high statistics using intense primary beam available at SIS in GSI and a thick production target.

\section{Simulation of Inclusive Spectra}
\label{simu}

\begin{figure}[h]
  \includegraphics[scale=0.237]{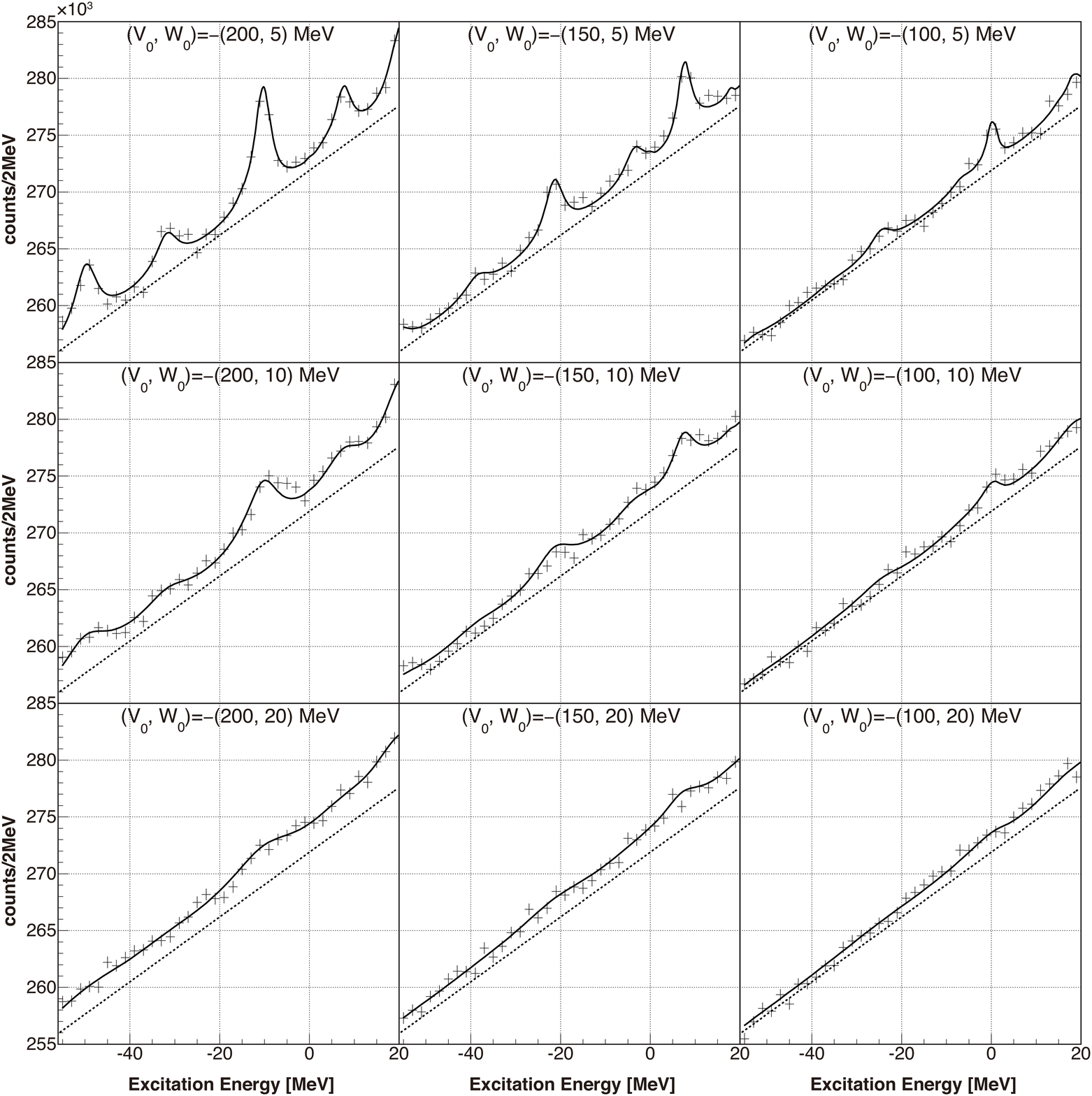}
\caption{Simulated inclusive spectra expected in 4.5-day data acquisition. 
$V_0$ is the real part and $W_0$ is the imaginary part of the optical potential at normal nuclear density.
In-medium mass reduction and width correspond to $|V_0|$ and $2|W_0|$, respectively. 
The amount of the background processes is shown by the dashed line.}
\label{fig_inclusive_spectra}       
\end{figure}

We have simulated inclusive spectra of the $^{12}$C($p$,$d$) reaction to discuss 
the experimental feasibility of finding peak structures. 
At first, we estimated the cross section of background processes in the inclusive ($p$,$d$) spectrum, which is mainly dominated by multi-pion production, 
based on COSY/ANKE data and simulation \cite{Ref9}. Then, combining with the formation cross sections calculated in Ref.~\cite{Ref10}, 
we simulated inclusive spectra expected in 4.5-day data acquisition.
Figure \ref{fig_inclusive_spectra} shows the result for several 
cases of different in-medium mass reductions and widths.
When the mass reduction is large and the width is small, distinct peak structures 
can 
be observed.
However, with smaller mass reduction and broader width, signal-to-noise ratio becomes much worse, and therefore peak structures can not seen in this case.
When $|V_0|=150$ MeV, as predicted by the NJL model calculations \cite{Ref2,Ref3}, and $|W_0|$ is less than $12.5$ MeV, as reported by the CBELSA/TAPS experiment \cite{Ref6},
there is a large chance to observe peaks in the missing-mass spectrum experimentally.
 
\section{Summary} 
\label{summ}
We are planning an inclusive spectroscopy experiment of $\eta '$ mesic nuclei with the  $^{12}$C($p$,$d$) reaction at GSI
aiming to study in-medium properties of the $\eta '$ meson. 
This experiment will be performed at GSI with intense proton beam supplied by SIS and
FRS as a spectrometer with good resolution. 
A simulation of inclusive spectra shows that 
significant structures 
will 
be observed even in inclusive spectra if the mass reduction is sufficiently large and the decay width is small.
 
The preparation, such as R\&D of the detectors and the optics mode for FRS, is on-going. 
We expect a first pilot experiment will be performed in 2013-2014. 

%
%
%
%
\begin{acknowledgements}
This work is supported by the Grant-in-Aid for Scientic Research (No. 20002003,
No. 20540273, No. 22105510, No. 22740161, No. 24105705, No. 24105707, No.
24105712, and No. 24540274) in Japan. Financial support by HIC-for-FAIR is
highly appreciated. This work was done in part under the Yukawa International
Program for Quark-hadron Sciences (YIPQS).

\end{acknowledgements}
%
%

\end{document}